\documentclass[10pt,showpacs,aps,prl,twocolumn,preprint,superscriptaddress]{revtex4}
\usepackage[dvips]{graphicx}
\usepackage{mathrsfs}
\usepackage{amsmath}
\usepackage{amssymb}
\usepackage{amsfonts}
\usepackage{ae}
\usepackage{units}

\begin{document}
%
\title{Cooling of a suspended nanowire by an AC Josephson current flow}
\author{Gustav Sonne}
\email{gustav.sonne@physics.gu.se}
\affiliation{Department of Physics, University of Gothenburg, SE-412 96 G\"oteborg, Sweden}
\author{Milton E. Pe\~na-Aza}
\affiliation{Department of Physics, University of Gothenburg, SE-412 96 G\"oteborg, Sweden}
\author{Leonid Y. Gorelik}
\affiliation{Department of Applied Physics, Chalmers University of Technology, SE-412 96 G\"oteborg, Sweden}
\author{Robert I. Shekhter}
\affiliation{Department of Physics, University of Gothenburg, SE-412 96 G\"oteborg, Sweden}
\author{Mats Jonson}
\affiliation{Department of Physics, University of Gothenburg, SE-412 96 G\"oteborg, Sweden}
\affiliation{School of Engineering and Physical Sciences, Heriot-Watt University, Edinburgh EH14 4AS, Scotland, UK}
\affiliation{School of Physics, Division of Quantum Phases and Devices, Konkuk University, Seoul 143-107, Korea}
\date{\today}
%
\begin{abstract}
We consider a nanoelectromechanical Josephson junction, where a
suspended nanowire serves as a superconducting weak link, and show
that an applied DC bias voltage can result in suppression of the
flexural vibrations of the wire. This cooling effect is achieved
through the transfer of vibronic energy quanta first to voltage driven
Andreev states and then to extended quasiparticle electronic
states. Our analysis, which is performed for a nanowire in the form of
a metallic carbon nanotube and in the framework of the density matrix
formalism, shows that such self-cooling is possible down to the
ground state of the flexural vibration mode of the nanowire.
\end{abstract}
%
\maketitle
Nanoelectromechanical systems (NEMS) have over the last two decades
been a very active field of both fundamental and applied
research. With typical dimensions on the nanometer scale NEMS combine
electronic and mechanical degrees of freedom for novel
applications. These include small mechanical resonators for
ultrasensitive mass spectroscopy \cite{Lassagne2008,Jensen2008} and
position sensing with close to quantum limited resolution
\cite{LaHaye,Etaki2008}. The prospect of a controlled fabrication of
NEMS devices whose physics is ultimately governed by the laws of
quantum mechanics is a major driving force in the field today.

Typically, nanoelectromechanical systems comprise a mechanical
resonator coupled to an electronic system used for both actuation and
detection. Due to the high mechanical vibration frequencies and the
exceptionally high quality factors recently achieved, these systems
allow for very low energy dissipation and extreme sensitivity to
external stimuli. To make full use of their potential, much research
has recently focused on the possibility to cool the mechanical
subsystem in NEMS to its vibrational ground state. Several theoretical
papers have invoked side-band cooling or dynamical back-action in
order to predict electromechanically induced cooling of mechanical
resonators to the ground state
\cite{Martin2004,Zippilli2009,WilsonRae2004,Ouyang2009}. The best
experimental result so far, corresponding to a vibron occupation
number of $3.8$, is due to Rocheleau {\it et al.}
\cite{Rocheleau2009}. They used an external electromagnetic field as
an energy transducer for side-band cooling of the mechanical
resonator. Similar results have been achieved by cooling with
dynamical back-action \cite{Naik}.

\begin{figure}
\includegraphics[width=0.4\textwidth]{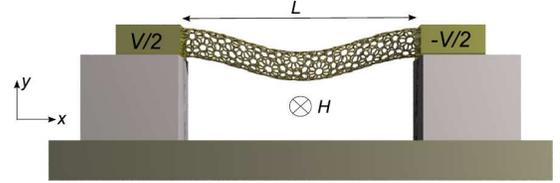}
\caption{(Color online) Schematic diagram of the Josephson junction
  discussed in the text. A suspended carbon nanotube of length $L$
  forms a weak link between two superconducting leads. An external
  magnetic field $H$ gives rise to a Lorentz force that couples
  flexural vibrations of the nanowire to the electrical current that
  is carried by Andreev states through the junction if it is biased by
  a voltage $V$. Cooling is achieved by the absorption of
  vibration-energy quanta as the electronic subsystem is excited
  across a gap in the Andreev spectrum (see Fig.~\ref{Andreevpic}).}
\label{picture}
\end{figure}

Here, we propose a new cooling method, which utilizes the unique
electronic properties of a Josephson junction. We consider an SNS
junction which is composed of a nanowire in the form of a metallic
carbon nanotube suspended between two superconducting leads as in
Fig.~\ref{picture}. Below we demonstrate that such a nanomechanical
weak link has the capacity to self-cool, if biased by a DC voltage
$V$, through the transfer of energy from the flexural vibrations of
the nanowire to a bath of electronic quasiparticle
excitations. Recently, it has been experimentally shown that for
systems similar to the one considered here it is possible to cool
nanomechanical objects below the ambient temperature through the
transfer of mechanical energy to electronic quasiparticle states
\cite{Koppinen2009,Muhonen2009}. Similar to these works, the cooling
process to be proposed is an inherent property of the system, which
makes the suggested mechanism a promising candidate for truly quantum
mechanical manipulation of mechanical resonators.

\begin{figure}
\includegraphics[width=0.47\textwidth]{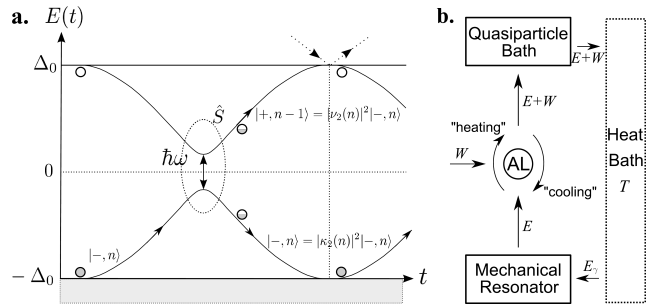} 
\caption{{\bf a.} Temporal evolution of voltage-driven Andreev levels
  (full lines) in the Josephson junction of Fig.~\ref{picture}. Here,
  $\vert\pm,n\rangle$ are states of the junction; $\vert\pm\rangle$
  denotes the upper/lower Andreev level and $\vert n\rangle$ is the
  quantum state of the nanowire. Transitions between $\vert
  -,n\rangle$ and $\vert +,n-1\rangle$, through the
  emission/absorption of one vibrational quantum, can be modeled by a
  scattering matrix $\hat{S}$. The transition probabilities depend on
  the state of the nanowire through
  $\vert\nu_2(n)\vert^2=1-\vert\kappa_2(n)\vert^2$. After one period,
  $T_V$, the partially filled Andreev levels join the continuum,
  creating quasiparticle excitations (represented by dashed arrows)
  and the electronic states are reset (filled and empty
  circles). Cooling of the nanowire occurs because absorption of
  vibrational quanta is more probable than emission at the ambient
  temperature. {\bf b.} Corresponding energy flow diagram. In each
  cycle the voltage-driven Andreev levels (AL) move some energy $E$
  from the mechanical resonator to a quasiparticle bath in thermal
  contact with a heat bath at temperature $T$. In this process the
  battery supplies an energy $W$. Only if the electronic and
  mechanical sub-systems are coupled can $E$ and $W$ be non-zero. The
  resonator is coupled by dissipation to the same heat bath from which
  it receives an energy $E_\gamma(=E$ in the steady
  state). ``Cooling'' [``heating''] refers to decreasing [increasing]
  the energy separation between the Andreev levels.}
\label{Andreevpic}
\end{figure}

The electromechanical coupling required for our cooling mechanism can
be achieved by applying an external magnetic field perpendicular to
the length of the nanowire. Displacement of the nanowire then couples
the mechanical and electronic degrees of freedom through the Lorentz
force. For an SNS junction that is short compared to the coherence
length in the superconductors, the AC Josephson dynamics induced by
the weak DC bias voltage can be expressed in terms of a pair of
Andreev states, which is periodically created and destroyed in the
junction. Between creation and destruction the Andreev levels evolve
adiabatically within the energy gap of the superconductors; the level
spacing first shrinks, reaches a minimum and is then restored to its
initial value as the Andreev states are dissolved in the continuum
quasiparticle spectrum (see Fig.~\ref{Andreevpic}). Being thermally
populated when created, the population of the Andreev levels become
``over-cooled'' as they evolve towards the middle of the gap. Since
the Andreev states are coupled to the nanowire vibrations, they can
absorb a quantum of vibration and approach an equilibrium distribution
through an electronic excitation from the lower to the upper Andreev
level. The absorbed energy is transferred to the electronic
quasiparticle continuum (and therefore transported away from the
junction) when the Andreev states are dissolved. Cooling of the
nanowire proceeds by repeated absorptions of vibrational quanta in
subsequent cycles of the temporal evolution of the Andreev levels.

To discuss the cooling mechanism quantitatively we consider the model
Hamiltonian,
\begin{align}
\hat{H}(t)&=\mathcal{E}(\phi(t))\hat{\sigma}_z+\Delta_0\sqrt{R}\sin(\phi(t)/2)\hat{\sigma}_x+\notag\\
&\hbar\omega\hat{b}^{\dagger}\hat{b}+\frac{2e}{\hbar}\frac{\partial \mathcal{E}(\phi(t))}{\partial \phi}LHy_0(\hat{b}+\hat{b}^{\dagger})\hat{\sigma}_z\,.
\label{Hamil}
\end{align}
Here, the first two terms describe the adiabatic dynamics
($\hbar\dot{\phi}\ll \Delta_0$) of the two level system formed by the
Andreev states that carry the Josephson supercurrent through the
(normal) nanowire \cite{Bagwell,Beenakker1991}; $\hat{\sigma}_\alpha$
are Pauli matrices, $\Delta_0$ is the order parameter in the
superconducting leads, $R$ is the normal state electronic reflection
probability, and $\mathcal{E}(\phi(t))=\Delta_0\cos(\phi(t)/2)$ is the
energy of the Andreev states for the completely transparent junction
with $\phi(t)$ the phase difference between the two
superconductors. The latter depends on the bias voltage according to
the Josephson relation $\dot{\phi}=2eV/\hbar$. The third term in the
Hamiltonian models the vibrating nanowire as a harmonic oscillator,
taking only the fundamental bending mode into account; the operator
$\hat{b}^{\dagger}$ $[\hat{b}]$ creates [annihilates] one quantum
$\hbar\omega$ of vibrational energy. The last term in \eqref{Hamil}
describes the electromechanical coupling of the current-carrying
Andreev states to the motion of the nanowire through the Lorentz
force. Here, $e$ is the electronic charge, $L$ is the length of the
wire, $y_0$ is the zero-point oscillation amplitude, $H$ is the
magnetic field and $2e/\hbar(\partial \mathcal{E}/\partial \phi)
\hat{\sigma}_z$ is the current operator in the nanowire
\cite{Gorelik1995}.

In the adiabatic regime, i.e. when $eV\leq eV_c=4R\Delta_0$, it is
convenient to use a basis set formed by the states
$\psi_{\pm}(\phi(t))$ with corresponding energies $E_{\pm}(t)=\pm
E(t)$, where $E(t)=\Delta_0(1-D\sin(\phi(t)/2))^{1/2}$ and $D=1-R$
\cite{footnote1}. In this basis, the Hamiltonian \eqref{Hamil} reads,
\begin{align}
\label{finalH}
\hat{H}_{eff}(t)=E(t)\hat{\tau}_z+\hbar\omega\hat{b}^{\dagger}\hat{b}+\Delta_0\Phi(\hat{b}^{\dagger}+\hat{b})\hat{\tau}_x\,.
\end{align}
Here, the Pauli matrices $\hat{\tau}_i$ span the space formed by the
states $\psi_{\pm}(\phi(t))$ (see Fig.~\ref{Andreevpic}). In the last
term of \eqref{finalH}, $\Phi=2LH\pi y_0/\Phi_0$ is the dimensionless
magnetic flux threading the area swept by the vibrating nanowire;
$\Phi_0=h/2e$ is the magnetic flux quantum. 

One notes that when $\phi=\pi$, the Andreev states
$\psi_{\pm}(\phi=\pi)$, with energies
$E_{\pm}(\phi=\pi)=\pm\Delta_0R^{1/2}$, are symmetric/antisymmetric
superpositions of states carrying current in opposite
directions. Transitions between the Andreev states --- induced by the
nanowire as it vibrates in a transverse magnetic field --- are
therefore by far most probable when $\phi(t=t_0)=\pi$. Below we will
therefore consider the resonant situation when
$\hbar\omega=2\Delta_0R^{1/2}$ \cite{footnote2}, which is the optimum
condition for the proposed cooling mechanism. To proceed further we
apply the rotating wave approximation, after which the Hamiltonian
reads
\begin{gather}
\label{Hfinal}
\hat{\mathscr{H}}_{eff}(t)=\begin{pmatrix}E(t)-\hbar\omega/2 & \Delta_0\Phi \hat{b}\\
 \Delta_0\Phi \hat{b}^{\dagger} & -E(t)+\hbar\omega/2\end{pmatrix}\,.
\end{gather}

Within the conditions for adiabaticity outlined above, we can now
analyze the probability for mechanically induced transitions between
the Andreev levels as a function of the coupling strength.  As such
transitions occur only in the vicinity of $t_0$, one can use the
parabolic expansion of $E(t)$ around this point, which leads to the
following time-dependent Schr\"{o}dinger equation,
\begin{gather}
\label{probampl}
i\partial_{\tau}\vec{c}_n(\tau)=\left(\tau^2\sigma_z+\Gamma\sqrt{n}\sigma_x\right)\vec{c}_n(\tau)\\
\tau=(t-t_0)(\xi/\hbar)^{1/3}\qquad \Gamma=\frac{\Phi\Delta_0}{\hbar\omega}\left(\frac{V_c}{V}\right)^{2/3}\notag\\
\xi=\frac{\partial^2E(t)}{\partial t^2}\bigg\vert_{t_0}=\frac{D(\hbar\omega)^3}{\hbar^2}\left(\frac{V}{V_c}\right)^2\,.\notag
\end{gather}
Here, $\vec{c}_n(\tau)=(c_{+,n-1}(\tau),c_{-,n}(\tau))^{T}$ where
$c_{\pm,n}(\tau)$ is the probability amplitude for finding the system
in the upper/lower electronic branch with the harmonic oscillator in
the state $n$.

Within the adiabatic regime, where $V\lesssim V_c$, the coupling
between the Andreev levels is relatively weak;
$\Gamma=\Delta_0\Phi/(\hbar\omega)\ll 1$ (see below). This allows us
to analyze \eqref{probampl} by treating $\Gamma$ as a perturbation to
the solution $\vert c_{-,n}(\tau)\vert^2=1$, where the system starts
in the lower electronic branch at time $-\delta\tau=-\delta
t(\xi/\hbar)^{1/3}$ and remains there until $+\delta\tau$. Solving
\eqref{probampl} for the probability $\vert c_{+,n-1}(\delta
\tau)\vert^2$ to find the system in the upper branch after passing
through the resonance, we find that $\vert
c_{+,n-1}(\delta\tau)\vert^2\simeq\pi n\Gamma^2$ which is consistent
with our numerical analysis.

In order to model the full evolution of the coupled electromechanical
system we evaluate the total density matrix $\hat{\rho}$ of the system
over one period. Under the assumption that the external temperature,
$T$, is much smaller than the superconducting gap, the system will
initially be found in the lower electronic branch with $n$ quanta of
the nanowire vibrations excited with probability $P_n^{in}$,
\begin{equation}
\hat{\rho}^{in}=\sum_{i,j=\pm}\sum_{n=0}^{\infty}P_n^{in}\vert i,n\rangle\langle j,n\vert=\sum_{n=0}^{\infty}P_n^{in}\begin{pmatrix}0 &0\\0 &1\end{pmatrix}\vert n\rangle \langle n\vert\notag\,.
\end{equation}
During the adiabatic evolution (no coupling between branches) the
system interacts with the external heat bath and the rate of change of
the density matrix is given by the expression,
\begin{align}
\partial\hat{\rho}(t)/\partial t=-i[\hat{\mathscr{H}}_{eff}(t)&,\hat{\rho}(t)]/\hbar+\gamma\hat{\mathfrak{L}}(\hat{\rho}(t))/2\\
\hat{\mathfrak{L}}(\hat{\rho})=-(1+n_B)&\left(\hat{b}^{\dagger}\hat{b}\hat{\rho}+\hat{\rho}\hat{b}^{\dagger}\hat{b}-2\hat{b}\hat{\rho}\hat{b}^{\dagger}\right)-\notag\\
&n_B\left(\hat{b}\hat{b}^{\dagger}\hat{\rho}+\hat{\rho}\hat{b}\hat{b}^{\dagger}-2\hat{b}^{\dagger}\hat{\rho}\hat{b}\right)\,,
\end{align}
where the collision integral $\hat{\mathfrak{L}}(\hat{\rho})$ models
the interaction of the mechanical subsystem with the environment.
Here, $n_B=(\exp(\beta\hbar\omega)-1)^{-1}$ with $\beta=(k_BT)^{-1}$
while $\gamma=\omega/Q$ is the damping rate of the nanowire vibrations
with $Q$ the quality factor. Considering the coupling to the
environment to be small ($Q\sim$\unit[$10^{5}$]{} \cite{Huttel2009})
we can proceed by treating the interaction of the system with the
external environment as a perturbation.

The large separation of the energy scales ensures that transitions
between the Andreev levels is almost always negligible. It is only
during a short time of order $\delta
t\simeq(\hbar\omega\Phi/\Delta_0)^{1/2}\hbar/( eV)\ll \pi\hbar/(eV)$
around $t_0$ that the two branches interact, which permits us to
describe the evolution of $\hat{\rho}$ through the resonance,
$\hat{\rho}(t_0+\delta t)=\hat{S}\hat{\rho}(t_0-\delta
t)\hat{S}^{\dagger}$, by the unitary scattering matrix $\hat{S}$,
\begin{gather}
\label{Smatrix}
\hat{S}=\begin{pmatrix}\kappa_1(\hat{n})& i\frac{\nu_1(\hat{n})}{\sqrt{\hat{n}+1}}\hat{b}\\ i\hat{b}^{\dagger}\frac{\nu_2(\hat{n}+1)}{\sqrt{\hat{n}+1}} &\kappa_2(\hat{n})\end{pmatrix}\,.
\end{gather}
In \eqref{Smatrix}, $\hat{n}=\hat{b}^{\dagger}\hat{b}$ is the vibron
number operator and the subscripts $1,2$ refer to the top/bottom
Andreev level respectively. The coefficient $\kappa_i(\hat{n})$
[$\nu_i(\hat{n})$] is the probability amplitude for the system to stay
in [scatter out of] the initial state $i$, which depends on the state
of the oscillator as outlined above
($\vert\kappa_i(n)\vert^2+\vert\nu_i(n)\vert^2=1$). As such,
$\vert\nu_2(n)\vert^2$ is the probability of the system, initially in
the lower electronic branch with $n$ vibrons excited, to scatter into
the upper electronic branch through the absorption of a vibron. It
thus corresponds to $\vert c_{+,n-1}(\delta \tau)\vert^2$ in
\eqref{probampl}. We conclude that $\vert\nu_2(n)\vert^2\simeq \pi
n\Gamma^2$ and note that
$\vert\nu_2(n)\vert^2=\vert\nu_1(n-1)\vert^2$, from the symmetry of
\eqref{probampl}.

After one period the bound Andreev states merge with the continuum of
itinerant states. At this time the quasiparticle carried by the
Andreev states is released into the continuum and the initial
conditions of the Andreev level populations is reset (see
Fig.~\ref{Andreevpic}) \cite{footnote3}. The mechanical system will,
however, not return to the initial vibrational state. Instead we must
find its final state from the density matrix for the mechanical system
after one period, $\hat{\rho}_{mech}^{f}$, obtained by tracing out the
electronic degrees of freedom in the total density matrix,
\begin{align}
\label{rhoevol}
\hat{\rho}_{mech}^{f}=\textrm{Tr}_{el}&(\hat{\rho}^{f})=\textrm{Tr}_{el}\bigg(\hat{S}\hat{\rho}^{in}\hat{S}^{\dagger}+\notag\\
\frac{\gamma}{2}&\frac{T_V}{2}\hat{S}\hat{\mathfrak{L}}(\hat{\rho}^{in})\hat{S}^{\dagger}+\frac{\gamma}{2}\frac{T_V}{2}\hat{\mathfrak{L}}(\hat{S}\hat{\rho}^{in}\hat{S}^{\dagger})\bigg)\,.\end{align}
\begin{figure}
\includegraphics[width=0.4\textwidth]{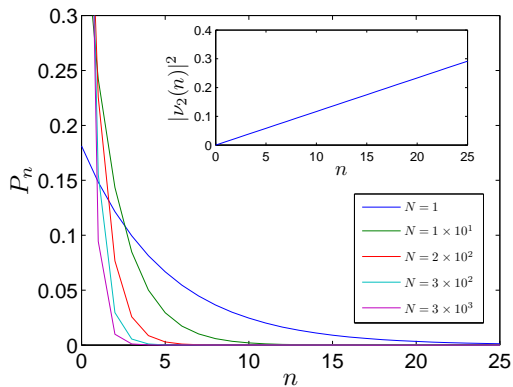}                
\caption{(Color online) Distributions of the probabilities $P_n$ for
  $n$ quanta of the lowest flexural vibration mode of the suspended
  nanowire in Fig.~\ref{picture} to be excited after $N$ of the
  cooling cycles described in Fig.~\ref{Andreevpic}. Initially $P_n$
  is thermally distributed, $P_n= \exp(-\beta\hbar\omega
  n)(1-\exp(-\beta\hbar\omega))$, with $k_BT=5\hbar\omega$. The plots
  were calculated for $\hbar\omega=$\unit[$10^{-6}$]{eV},
  $\Delta_0=10\hbar\omega$, $y_0=$\unit[20]{pm}, $L=$\unit[100]{nm},
  $H=$\unit[1]{T} and $T_V\sim$\unit[20]{ns}. The inset shows
  $\vert\nu_2(n)\vert^2$ for the same parameters.}
\label{densitypic}
\end{figure}

Equation \eqref{rhoevol} describes the evolution of the mechanical
density matrix over one period. Without coupling to the environment,
$\gamma=0$, this corresponds to a decay of the mechanical subsystem,
since for each period there is a probability $\vert\nu_2(n)\vert^2$
that the state $\vert -,n\rangle$ scatters into the state $\vert
+,n-1\rangle$. As the opposite process is forbidden if initially only
the lower electronic branch is populated, the mechanical subsystem
would thus approach the vibrational ground state at a rate, which
depends on the strength of the coupling. Including effects of the
environment, however, the mechanical subsystem will not reach its
ground state as the external damping drives the system towards thermal
equilibrium.  In Fig.~\ref{densitypic} we plot the evolution --- over
many periods starting from its initial thermal distribution --- of the
probability $P_n$ for $n$ quanta of mechanical nanowire vibrations to
be excited using realistic experimental parameters. As can clearly be
seen, the scattering process acts to lower the number of excited
quanta, effectively cooling the nanowire down to a final average
vibron population $\langle n\rangle \sim 0.1$. This is confirmed by a
perturbative calculation with $\Gamma\ll 1$, which shows that to first
order in the damping rate, $\gamma$, the stationary population of the
first excited state is
\begin{equation}
P_1^{stat}\simeq \frac{1}{Q\Phi^2}\frac{\hbar\omega}{\Delta_0}\left(\frac{V}{V_c}\right)^{1/3}\frac{1}{e^{\beta\hbar\omega}-1}\sim 0.1\,,
\end{equation}
which is in accordance with the distribution in
Fig.~\ref{densitypic}. Here we note that including a small initial
population of the upper Andreev branch, $c_+^{in}\ll 1$, only
changes the stationary vibron population to $\sim \langle n\rangle
+O(c_+^{in})$. Thus, including, e.g., a thermal population of the
upper Andreev branch only increases the stationary vibron population
by a factor $c_+^{in}\simeq \exp(-2\beta\Delta_0)\sim 0.02$, and
does not change the qualitative results of this paper.

To conclude, we have shown that quantum mechanical cooling of a
nanomechanical resonator acting as the weak link in a Josephson
junction is possible. In particular, we have considered the example of
a suspended carbon nanotube, where the unique combination of high
resonance frequencies and high mechanical quality factors together
with a high transparency for electrons combine to give sufficiently
strong coupling for efficient cooling. Using realistic experimental
parameters we have shown that for a short suspended nanotube, a
stationary distribution of mechanical vibrational quanta corresponding
to an average occupation number as low as $\langle n\rangle =0.1$ can
be achieved. This is truly in the quantum mechanical
regime. Furthermore, the suggested mechanism does not rely on any
external electromagnetic fields to stimulate the cooling
process. Rather, the proposed system should act as a self-cooling
device as the ``over-cooled'' Andreev states can readily absorb energy
from the mechanical subsystem given that sufficient coupling between
the two can be achieved. The corresponding energy uptake of the
electronic subsystem is later released into the quasiparticle
continuum, leading to an effective cooling of the nanomechanical
resonator.

L. Y. Gorelik thanks V. S. Shumeiko for fruitful discussions. This
work was supported in part by the Swedish VR and SSF, by the EC
project QNEMS (FP7-ICT-233952) and by the Korean WCU program
(MEST/KOSEF R31-2008-000-10057-0).
\bibliography{Paper_update}

\begin{thebibliography}{10}

\bibitem{Lassagne2008}
B.~Lassagne {\em et~al.},
\newblock Nano Lett. {\bf 8}, 3735 (2008).

\bibitem{Jensen2008}
K.~Jensen, K.~Kim, and A.~Zettl,
\newblock Nat. Nanotechnol. {\bf 3}, 533 (2008).

\bibitem{LaHaye}
M.~D. LaHaye {\em et~al.},
\newblock Science {\bf 304}, 74 (2004).

\bibitem{Etaki2008}
S.~Etaki {\em et~al.},
\newblock Nat. Phys. {\bf 4}, 785 (2008).

\bibitem{Martin2004}
I.~Martin {\em et al.},
\newblock Phys. Rev. B {\bf 69}, 125339 (2004).

\bibitem{Zippilli2009}
S.~Zippilli, G.~Morigi, and A.~Bachtold,
\newblock Phys. Rev. Lett. {\bf 102}, 096804 (2009).

\bibitem{WilsonRae2004}
I.~Wilson-Rae, P.~Zoller, and A.~Imamo\={g}lu,
\newblock Phys. Rev. Lett. {\bf 92}, 075507 (2004).

\bibitem{Ouyang2009}
S.~H. Ouyang, J.~Q. You, and F.~Nori,
\newblock Phys. Rev. B {\bf 79}, 075304 (2009).

\bibitem{Rocheleau2009}
T.~Rocheleau {\em et~al.},
\newblock Nature {\bf 463}, 72 (2010).

\bibitem{Naik}
A.~Naik {\em et~al.},
\newblock Nature {\bf 443}, 193 (2006).

\bibitem{Koppinen2009}
P.~J. Koppinen and I.~J. Maasilta,
\newblock Phys. Rev. Lett. {\bf 102}, 165502 (2009).

\bibitem{Muhonen2009}
J.~T. Muhonen {\em et~al.},
\newblock Appl. Phys. Lett. {\bf 94}, 073101 (2009).

\bibitem{Bagwell}
P.~F. Bagwell,
\newblock Phys. Rev. B {\bf 46}, 12573 (1992).

\bibitem{Beenakker1991}
C.~W.~J. Beenakker,
\newblock Phys. Rev. Lett. {\bf 67}, 3836 (1991).

\bibitem{Gorelik1995}
L.~Y. Gorelik {\em et~al.},
\newblock Phys. Rev. Lett. {\bf 75}, 1162 (1995).

\bibitem{footnote1}
In this regime, the voltage-induced Landau-Zener transition at
$\phi=\pi$ between the adiabatic Andreev levels is suppressed,
ensuring that without electromechanical coupling the level population
remains constant in time.

\bibitem{footnote2}
This condition can be achieved by controlling the magnitude of the
order parameter $\Delta_0$ through the magnetic field for junctions
with small reflection coefficients \cite{Kong2001}.

\bibitem{Kong2001}
J.~Kong {\em et~al.},
\newblock Phys. Rev. Lett. {\bf 87}, 106801 (2001).

\bibitem{Huttel2009}
A.~K. H\"{u}ttel {\em et~al.},
\newblock Nano Lett. {\bf 9}, 2547 (2009).

\bibitem{footnote3}
As explained in \cite{Gorelik1998}, the non-equilibrium population of
the Andreev states that approach the edges of the quasiparticle energy
spectrum at $E=\pm \Delta_0$ (Fig.~\ref{Andreevpic}) cannot, due to
selection rules, affect the population of the new Andreev states,
which are created with populations ``reset'' to the thermal
equilibrium and move away from the edges. This system will not be
heated since (i) the released energy is carried away by quasiparticles
\cite{Bratus1997}, which escape from the vicinity of the weak link and
spread over the reservoirs with greatly reduced density and (ii) the
``escape time'', $T_{\textrm{esc}}\sim(eV/\Delta_0)^{1/3}T_V$, is
short compared to the period $T_V$.

\bibitem{Gorelik1998}
L.~Y. Gorelik {\em et~al.},
\newblock Phys. Rev. Lett. {\bf 81}, 2538 (1998).

\bibitem{Bratus1997}
E.~N. Bratus {\em et~al.},
\newblock Phys. Rev. B {\bf 55}, 12666 (1997).

\end{thebibliography}
\end{document}